**Visualizing the Formation of the Kondo Lattice and the Hidden Order in URu$_2$Si$_2$**

Pegor Aynajian[1][§], Eduardo H. da Silva Neto[1][§], Colin V. Parker[1][§], Yingkai Huang[2], Abhay Pasupathy[3], John Mydosh[4], Ali Yazdani[1*]

[1]Joseph Henry Laboratories and Department of Physics, Princeton University, Princeton, NJ 08544, USA

[2]van der Waals-Zeeman Institute, University of Amsterdam, 1018XE Amsterdam, The Netherlands

[3]Department of Physics, Columbia University, New York, NY 10027, USA

[4]Kamerlingh Onnes Laboratory, Leiden University, 2300 RA Leiden, The Netherlands

*To whom correspondence should be addressed, email: yazdani@princeton.edu

[§]These authors contributed equally to this work

**Heavy electronic states originating from the _f_ atomic orbitals underlie a rich variety of quantum phases of matter (1-5). We use atomic scale imaging and spectroscopy with the scanning tunneling microscope (STM) to examine the novel electronic states that emerge from the uranium _f_ states in URu$_2$Si$_2$. We find that as the temperature is lowered, partial screening of the _f_ electrons' spins gives rise to a spatially modulated Kondo-Fano resonance that is maximal between the surface U atoms. At T=17.5 K, URu$_2$Si$_2$ is known to undergo a 2$^{nd}$ order phase transition from the Kondo lattice state into a phase with a hidden order parameter (6-9). From tunneling spectroscopy, we identify a spatially modulated, bias-asymmetric energy gap with a mean-field temperature dependence that develops in the hidden order state. Spectroscopic imaging further reveals a spatial** **correlation between the hidden order gap and the Kondo resonance, suggesting that the two phenomena involve the same electronic states.**

A remarkable variety of collective electronic phenomena have been discovered in compounds with partially filled $f$ orbitals, where electronic excitations act as heavy fermions (1, 2) . Like other correlated electronic systems, such as the high temperature superconducting cuprates, several of the heavy fermion compounds display an interplay between magnetism and superconductivity, and have a propensity toward superconducting pairing with unconventional symmetry (1-5). However, unlike cuprates, or the newly discovered ferropnictides, the heavy fermion systems do not suffer from inherent dopant-induced disorder and offer a clean material system for the study of correlated electrons. The local $f$ electrons interact both with the itinerant *spd* electrons as well as with each other, resulting in a rich variety of electronic phases. In many of these materials, screening of the local moments by the Kondo effect begins at relatively high temperatures resulting in heavy fermion state at low temperatures. Exchange interactions between the local moments become more important at lower temperatures and can result in the formation of magnetic phases as well as superconductivity at even lower temperatures. Among the heavy fermion compounds perhaps the most enigmatic is the $URu_2Si_2$ system, which undergoes a $2^{nd}$ order phase transition with a rather large change in entropy (6-8) at 17.5 K from a paramagnetic phase with Kondo screening to a phase with an unknown order parameter (9). This material possesses low-energy commensurate and incommensurate spin excitations, which are gapped below the hidden order (HO) transition temperature (10-13). These features are believed to be signatures of a more complex order parameter, the identification of which has so far not been possible despite numerous investigations (12-18). Moreover, analogous to other correlated systems, this unusual conducting phase is transformed into an unconventional superconducting state at 1.5 K (6, 8, 19), the understanding of which hinges on formulating the correct model of the hidden order state.

We report scanning tunneling microscopy (STM) measurements on $URu_2Si_2$ single crystals that allow atomic scale examination of this system in the paramagnetic Kondo phase and its phase transition into the hidden order (HO) state. We isolate electronic signatures of the Kondo lattice state and their transformation at the HO transition. Although there have long been reports on modification of the electronic structure of $URu_2Si_2$ at the onset of the HO transition, such as those in specific heat (6, 8) as well as optical (20, 21), point contact (21, 22), and angle resolved photoemission spectroscopy measurements (17, 23, 24), our experiments provide an unprecedented determination of these changes with high energy and spatial resolutions. We find a particle-hole asymmetric energy gap that turns on with a mean-field temperature dependence near the bulk HO transition. More importantly, spectroscopic mapping as a function of temperature further reveals that the hidden order gap is spatially correlated on the atomic scale with the electronic signatures of the Kondo lattice state.

We carried out our experiments in a specially designed STM using high quality single crystal $URu_2Si_2$ samples that were cleaved in situ in UHV prior to measurements. The STM topographies show that cleaved surfaces can be terminated with primarily two types of surfaces, one of which is atomically ordered with a lattice spacing corresponding to either the U or the Si layer (Figure 1A; termed surface (A)), while the other is reconstructed (Figure 1B; termed surface (B)) (see more information in supplementary section). The occurrence of surfaces (A) and (B) with roughly equal probabilities (55% and 45% respectively) implies that these surfaces are the two sides of the same cleave, suggesting that the cleaving process involves breaking of a single type of chemical bond. Moreover, the relative height between surfaces (A) and (B) shown in Fig. 1C, reveals that surface (A) lies ~1.5Å above and ~3.3Å below surface (B). This asymmetry allows us to uniquely identify surface (A) and (B) as the U and Si layers, respectively (see Fig.1E). Any other possibility (i.e, Fig.1F) will require the breaking of two bonds, the result of which would be the observation of four surfaces with 25% probabilities. Occasionally (<5% of the time) small patches of a different surface (termed surface (C)) have also been observed (Fig.1D). The presence of multiple surfaces for cleaved $URu_2Si_2$ samples indicates that obtaining local surface structure information is critical to identifying which spectroscopic properties are most related to the bulk properties. This requirement puts other surface sensitive spectroscopic techniques, such as angle resolved photoemission or point contact spectroscopy, at a disadvantage.

The temperature evolution of the spatially averaged STM spectra on the atomically ordered terraces (U terminated, surface (A)), shown in Fig. 2B, reveals the development of electronic correlations in $URu_2Si_2$ below 120 K, which evolve into dramatic spectroscopic signatures of the HO phase as the temperature is reduced (Fig. 2C-D). At high temperatures (above 120 K), the spectrum presents a broad feature that has weak energy dependence, although it shows a slightly asymmetric density of states (DOS) for electron or hole tunneling. As the temperature is lowered, we find that the background is modified with an asymmetric resonance near the Fermi level, which sharpens as the temperature is reduced. On further cooling below the HO transition, we find the opening of a low energy gap that widens with decreasing temperature and reveals an unusual asymmetry relative to the Fermi energy and an even sharper structure within the gap (Fig. 2D). The unusual shape of the low energy spectral properties and their dramatic evolution with temperature provide important clues to the underlying correlations responsible for the thermodynamic phases of $URu_2Si_2$.

Focusing on the spectroscopic features above the HO transition, we note that the asymmetric lineshape near the Fermi level developing below 120 K is reminiscent of a Fano spectral lineshape (25-28) measured for single Kondo impurities on the surface of noble metals (29-31). Recent calculations of the local electronic density of states on $URu_2Si_2$ reveal a Kondo resonance with a Fano lineshape on the U-layer in qualitative agreement with our data (18). Previous thermodynamic and transport studies have also identified the temperature range above the HO transition as being dominated by Kondo screening of the *f* electron moments (6, 8, 19). For STM experiments, the Fano lineshape naturally occurs because of the presence of two interfering tunneling paths from the STM tip, one directly into the itinerant electrons, and the other indirectly through the Kondo resonance. The Fano lineshape,

$$G(V) \propto \frac{((V-E_0)/\Gamma + q)^2}{1 + ((V-E_0)/\Gamma)^2},$$

(Eq. 1)

is characterized by the resonance energy $E_0$, its width $\Gamma$, and the ratio of probabilities between the two tunneling paths $q$. The addition of this lineshape to the high temperature background density of states accurately fits the spectroscopic data over a wide range of temperatures (red line in Fig. 2B, see also supplementary information). The extracted values of *q=1.3±0.3* and *E_0=5±2 meV* show no significant temperature dependence within the uncertainty of our fits above the HO transition, whereas $\Gamma$ clearly broadens with increasing temperature as shown in Fig. 3A. Results for the single channel spin one-half Kondo impurity model in a Fermi liquid regime have been used to describe the temperature dependence of $\Gamma$ (31),

$$\Gamma = 2\sqrt{(\pi k_B T)^2 + 2(k_B T_K)^2},$$

(Eq. 2)

and to extract the value of the Kondo temperature, $T_K$=129±10 K (Fig 2A). The success of this model at describing the spectra above the HO transition is somewhat surprising given that the spins associated with the *f* levels reside on a periodic U sub-lattice (32) and that the material displays non-Fermi liquid transport properties at low temperatures (33).

To obtain information about the influence of the periodic arrangement of Kondo "impurities" in our lattice, we now turn to the spatial dependence of the STM spectra. Obtaining topographies and STM differential conductance maps (dI/dV) with sub-Angstrom resolution we find that the DOS modulates with the same periodicity as the atomic lattice (Fig. 3C & D). Such a variation can be expected since the amplitudes for the two interfering paths that cause the Kondo-Fano resonance depend sensitively on the path that the electrons have to travel from the tip to the sample. Fitting individual spectra at each location with the Fano lineshape, we find that the extracted ratio $q$, which describes the relative strength for tunneling into the Kondo resonance and the $spd$ bands, shows measurable variation on the atomic scale (Fig. 3E). Naively, we would expect $q$ to be maximal when the tip is over the U atoms. However, the orientation and shape of the atomic orbitals also plays a role in the tunneling into the Kondo resonance (34, 35). In fact, the spatial dependence of $q$ shows enhanced tunneling into the resonance when the tip is over the locations between the surface U atoms. A schematic of the orientation of the U's f orbitals at a cleaved surface, shown in Fig. 3B, provides a possible way to understand such spatial dependence.

As illustrated by the spectra in Fig. 2, the development of the HO has a dramatic influence on the Kondo-Fano resonance and contains key information on the electronic correlations responsible for this transition. In order to isolate the spectroscopic changes at the onset of the HO, we divide the spectra obtained at temperatures below $T_{HO}$=17.5 K with the spectrum measured at 18 K just above the bulk hidden order transition, as shown in Fig. 4A and 4B (see supplementary information). The normalized spectra reveal the onset of a low energy gap centered at an energy below the Fermi level. While the presence of a gap associated with the HO has been indicated by previous thermodynamic (6, 8) and spectroscopic measurements (20-22, 36), our high-resolution data and their spatial dependence allows identification of several of its fundamental features.

A phenomenological way to characterize the HO gap, $\Delta_{HO}$, is to fit the spectra to a thermally convoluted DOS with a BCS function form, from which we can extract the magnitude and temperature dependence of $\Delta_{HO}(T)$. Following this procedure (Fig.4A-B) we extract $\Delta_{HO}(T)$, which evolves more rapidly than simple thermal broadening (see supplementary information). However, as we approach the transition temperature, $\Delta_{HO}$ becomes comparable to thermal broadening, making the precise determination of the onset temperature difficult. Regardless, we find the temperature dependence of $\Delta_{HO}(T)$ to follow a mean field behavior with onset temperature of $T_{HO}{\sim}16K$ (Fig. 4C). Broken symmetry at the surface is likely to influence the HO state and may account for the slightly reduced observed onset temperature relative to that of bulk measurements. An important aspect of the $\Delta_{HO}$ is the fact that it develops asymmetrically relative to the Fermi energy and it shifts continuously to lower energies upon lowering of the temperature (Fig. 2C-D). We quantify the changes to $\Delta_{HO}$ and its offset by fitting the data to a BCS function form with an offset energy relative to $E_F$ (Fig. 2C-D, Fig. 4D; see caption of Fig. 4).

The low temperature extrapolation, $\Delta_{HO}(0)$=4.1±0.2 meV, yields $2\Delta_{HO}(0)/k_B T_{HO}$=5.8±0.3, which together with the value of the specific heat coefficient $\gamma_c = C/T$ for $T > T_{HO}$ (8) within the BCS formalism results in a specific heat jump at the transition of $\Delta C$=6.0±1.3 J K$^{-1}$ mol$^{-1}$, consistent with previous measurements (7, 8, 12). The partial gapping of the Fermi surface observed in our spectra also corroborates the recently observed gapping of the incommensurate spin excitations by inelastic neutron scattering (INS) experiments (12). Finally, the spectrum develops additional, sharper features within $\Delta_{HO}$ at the lowest temperatures (Fig 4B). Such lower energy features may be related to the gapping of the commensurate spin excitations at the antiferromagnetic wavevector below $T_{HO}$ also seen in INS at an energy transfer of about 2 meV (11, 12, 13).

The spatial variation of the STM spectra provides additional information about the nature of redistribution of the electronic states that gives rise to $\Delta_{HO}$. Displayed in Fig. 5, we show energy-resolved spectroscopic maps measured above and below $T_{HO}$, all of which show modulation on the atomic scale. The measurements above $T_{HO}$ show no changes in their atomic contrast within the energy range where the $\Delta_{HO}$ is developed. In fact, the modulations in these maps (5B-E) are due to the surface atomic structure but occur with a contrast that is opposite to that of the STM topographies of the same region (Fig. 5A). However, observation of reverse contrast in STM conductance maps is expected as a consequence of the constant current condition. Similar measurements below $T_{HO}$ are also influenced by the constant current condition, as shown in Fig. 5G-J;

nonetheless, these maps show clear indication of the suppression of contrast associated with $\Delta_{HO}$ at low energies (within the gap, see Fig. 5F) and the consequent enhancement at high energies (just outside the gap).

To isolate the spatial structure associated with $\Delta_{HO}$ and to overcome any artifacts associated with the measurement settings, we divide the local conductance measured below $T_{HO}$ by that above for the same atomic region, as shown in Fig. 5L-O. Such maps for $|V| < \Delta_{HO}$ illustrate that the suppression of the spectral weight principally occurs in between the surface U atoms. These maps are essentially the spatial variation of the conductance ratios, shown in Fig. 4A. Therefore, consistent with the BCS-like redistribution of spectral weight, we find that conductance map ratios at energies just above $\Delta_{HO}$ illustrate an enhancement between the surface U atoms. Quantifying these spatial variations further, we also plot the correlation between the conductance map ratios and the atomic locations above and below $T_{HO}$ (Fig. 5K) to show that $\Delta_{HO}$ is *strongest* in between the surface U atoms, i.e., at the same sites where tunneling to the Kondo resonance is enhanced (Fig. 3E). Our observation that the modulation in the tunneling amplitude into the Kondo resonance correlates with the spatial structure of the hidden order gap shows that the two phenomena involve the same electronic states.

Our finding of an asymmetric mean field-like energy gap would naively suggest the formation of a periodic redistribution of charge and/or spin at the onset of the HO due to Fermi surface nesting. However, consistent with previous scattering experiments (8, 11, 12, 13), we find no evidence for any conventional density wave in our experiments. Recently it has been suggested that the opening of a gap may be a consequence of hexadecapole ordering emerging from the crystal field splitting of U's $5f^2$ states (18). While the experimentally observed gap can be the consequence of the hexadecapole ordering, no signatures of the higher temperature crystal field splitting was observed in our data. Other proposals indicate that the development of the gap is the effect of a hybridization gap (37, 38). Though a conventional hybridization gap should not manifest itself as a $2^{nd}$ order phase transition with a mean-field-like order parameter, an exotic form of hybridization accompanied by orbital ordering can be a possible candidate for the hidden order phase. Regardless, our spatial mapping, which reveals the enhancement of the Kondo-Fano signature between the surface U atoms and the relatively stronger hidden order gap at these same locations, provides a key to identifying the electronic states responsible for these phenomena, and therefore the understanding of the nature of the hidden order phase.

## Acknowledgments

We gratefully acknowledge discussions with P. W. Anderson, P. Chandra, P. Coleman, K. Haule, G. Kotliar, and A. Pushp. This work was supported by Department of Energy Basic Energy Sciences and in parts by the National Science Foundation and by W. M. Keck Foundation. P. A. acknowledges postdoctoral fellowship support through the Princeton Center for Complex Materials funded by National Science Foundation MRSEC program.

**Figure Legends**

**Figure 1. STM topography.** (**A, B**) Constant current topographic image (-200 mV, 60 pA, 33 Å) showing an atomically ordered surface (termed surface (A)) and (100 mV, 200 pA, 90 Å) showing an atomic layer with surface reconstruction (termed surface (B)), respectively. (**C**) The relative heights between the surfaces (A) and (B). (**D**) Constant current topographic image (-50 mV, 100 pA) over a 185x140 $Å^2$ area showing a (2x1) reconstructed surface (surface (C)) lying ~2.2Å above surface (A). A horizontal line-cut through the data in (**C & D**) is shown on the bottom panels. (**E**) Schematic diagram illustrating the different atomic layers of $URu_2Si_2$. U is identified as the atomically ordered surface (surface (A)) which lies 1.24Å above and 3.56Å below surface (B). In this case, obtaining surfaces (A) and (B) requires breaking of a single bond only (U-Si, see arrows). (**F**) Schematic diagram illustrating a different possibility for the cleaved surfaces, which requires the breaking of two bonds (Ru-Si and U-Si; see arrows). This cannot explain why surfaces (A) and (B) occur with roughly equal probabilities. The step heights in (**A**) and (**B**) are obtained (or calculated) from (6).

**Figure 2. STM topography and spectroscopy on $URu_2Si_2$.** (**A**) Constant current topographic image (-200 mV, 60 pA) over a 200 Å area showing the atomically ordered surface where the spectroscopic measurements are performed. (**B**) and (**C**) Averaged electronic density of states above (**B**) and below (**C**) the hidden order temperature. The red lines in (**B**), (**C**) are the results of least squares fit described in the text and supplementary information. Spectra are offset by 0.25 nS for clarity. (**D**) Averaged electronic density of states at low temperatures showing additional features within the gap. Spectra are offset by 1 nS for clarity.

**Figure 3. Kondo lattice.** (**A**) Temperature dependence of the Kondo resonance width $\Gamma$ extracted from the fits in Fig.2B. The red line represents the temperature dependence for a single Kondo impurity described in the text which results in a Kondo temperature $T_K$=129±10 K. (**B**) Crystal structure of $URu_2Si_2$ indicating the different atomic layers and a schematic of the orbitals which bond the Si atoms to the U atoms. (**C**) A high resolution constant current topography of 4x4 atoms taken at 18 K. (**D**) Conductance map at 6mV (Kondo resonance energy) corresponding to the topography in (**C**) showing atomic scale modulations. (**E**) The dimensionless q(r) map on the same area as in (**C**) obtained by fitting the spectra at each location to a Fano lineshape. The larger values of q (indicating higher tunneling probability to the Kondo resonance) lie in between the atomic sites as depicted by the black square.

**Figure 4. Temperature dependence of the hidden order gap.** (**A**),(**B**) The experimental data below $T_{HO}$ divided by the 18 K data. The data are fit to the form $D(V) = (V - V_0 - i\gamma) / \sqrt{(V - V_0 - i\gamma)^2 - \Delta^2}$, which resembles an asymmetric BCS-like density of states with an offset from $E_F$. $V_o$, $\gamma$, and $\Delta$ are the gap position (offset from the Fermi energy), the inverse quasi-particle lifetime, and the gap magnitude, respectively. A quasi-particle lifetime broadening of $\gamma$~1.5 mV was extracted from the fits. (**C**) Temperature dependence of the gap extracted from the fits in (**A**) (black squares) and from a direct fit to the raw data of Figure 2C (blue circles). Both results are comparable within the error bars. The transition temperature $T_{HO}=16.0\pm0.4K$ is slightly lower than the bulk transition temperature presumably as a consequence of the measurement being performed on the surface. (**D**) Temperature dependence of the gap position $V_0$ extracted from the fits. The line is a guide to the eye.

**Figure 5. Atomic origin of the hidden order.** (**A**) A constant current topography at $T$=18 K corresponding to 5x5 atoms. Spatial conductance maps on the area in (**A**) at different bias voltages $V$ obtained at 18 K (**B-E**) and 6.6 K (**G-J**). The junction is stabilized at -100 mV and 100 pA. All conductance maps are normalized to their mean value to emphasize the atomic contrast. The maps display an atomically periodic modulation. (**F**) Average dI/dV spectra at 18K and 6.6K. The arrows indicate the energies where the conductance maps are performed. (**L-O**) Division of the *G(r,V,T=6.6 K)* maps with the *G(r,V,T=18 K)* maps showing a contrast reversal of the conductance when moving from outside the gap (**L**;$V$=-6 mV) to within the gap (**M,N**). The loss of the spectral weight in the gap and the transfer to higher energies occurs principally between the surface U atoms as is shown by the white square boxes. (**K**) Correlation of the conductance maps with the atomic locations above and below the hidden order temperature showing a dramatic change of correlation (change of contrast) within the gapped region between the surface U atoms. Homogeneous gapping should result in no change of correlation.

**Fig. 1**

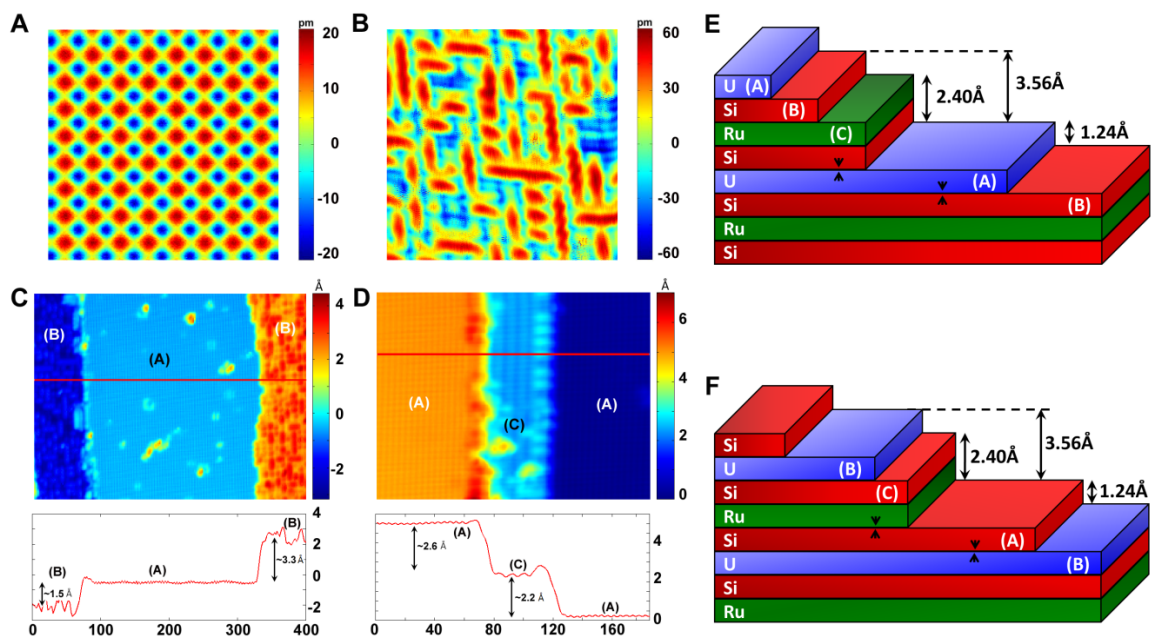

**Fig. 2**

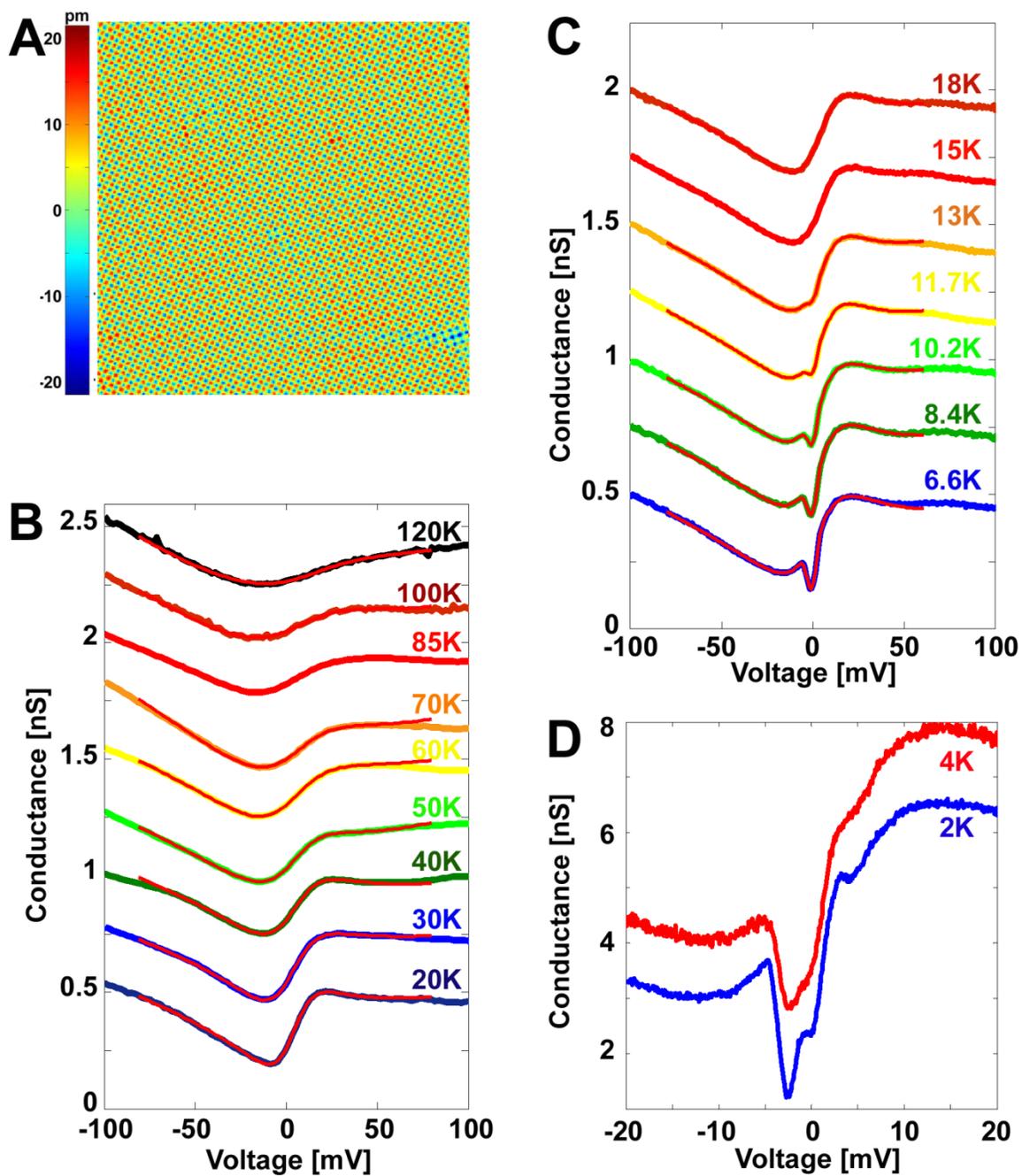

**Fig. 3**

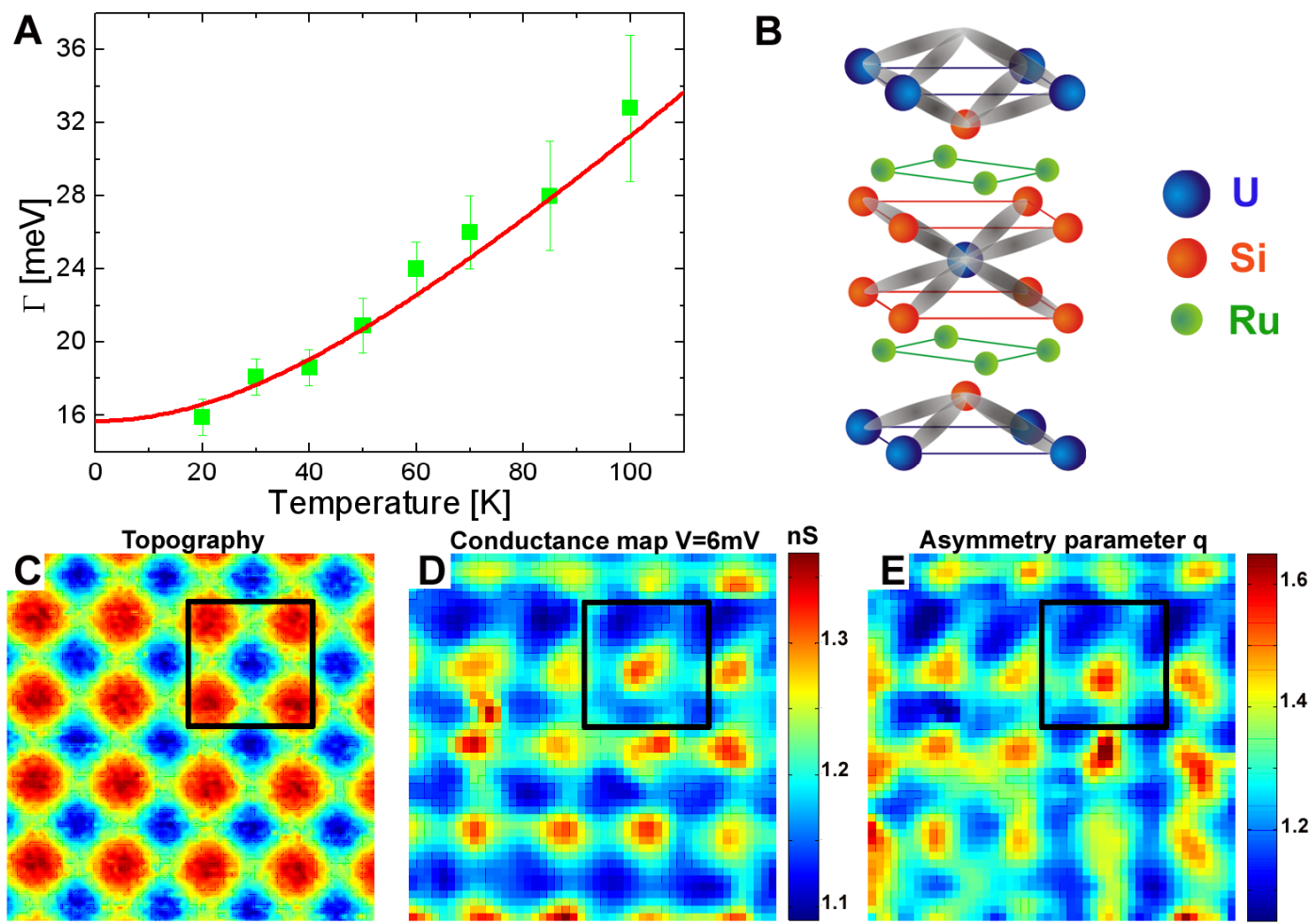

Topography

Conductance map V=6mV    nS

Asymmetry parameter q

**Fig. 4**

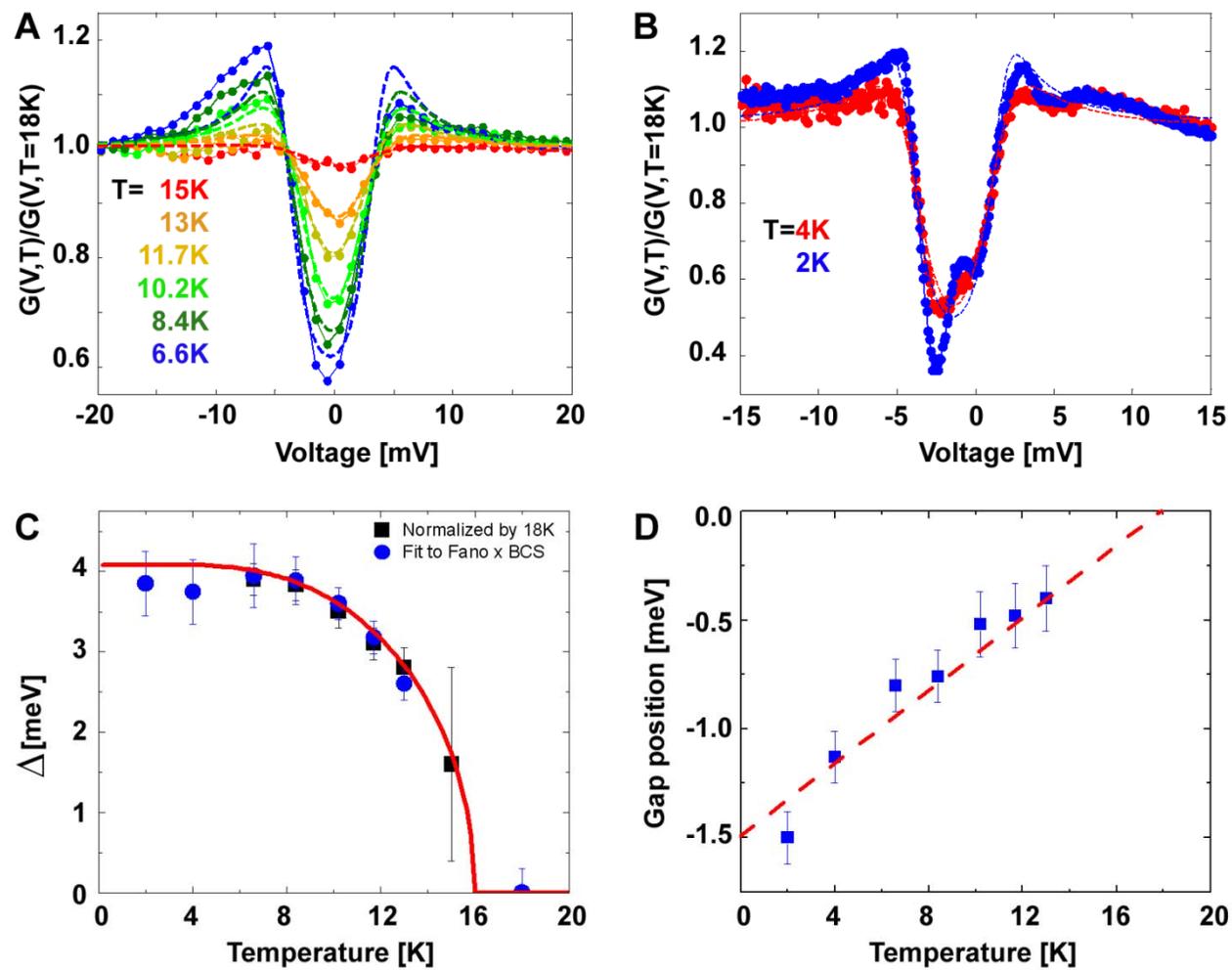

**Fig. 5**

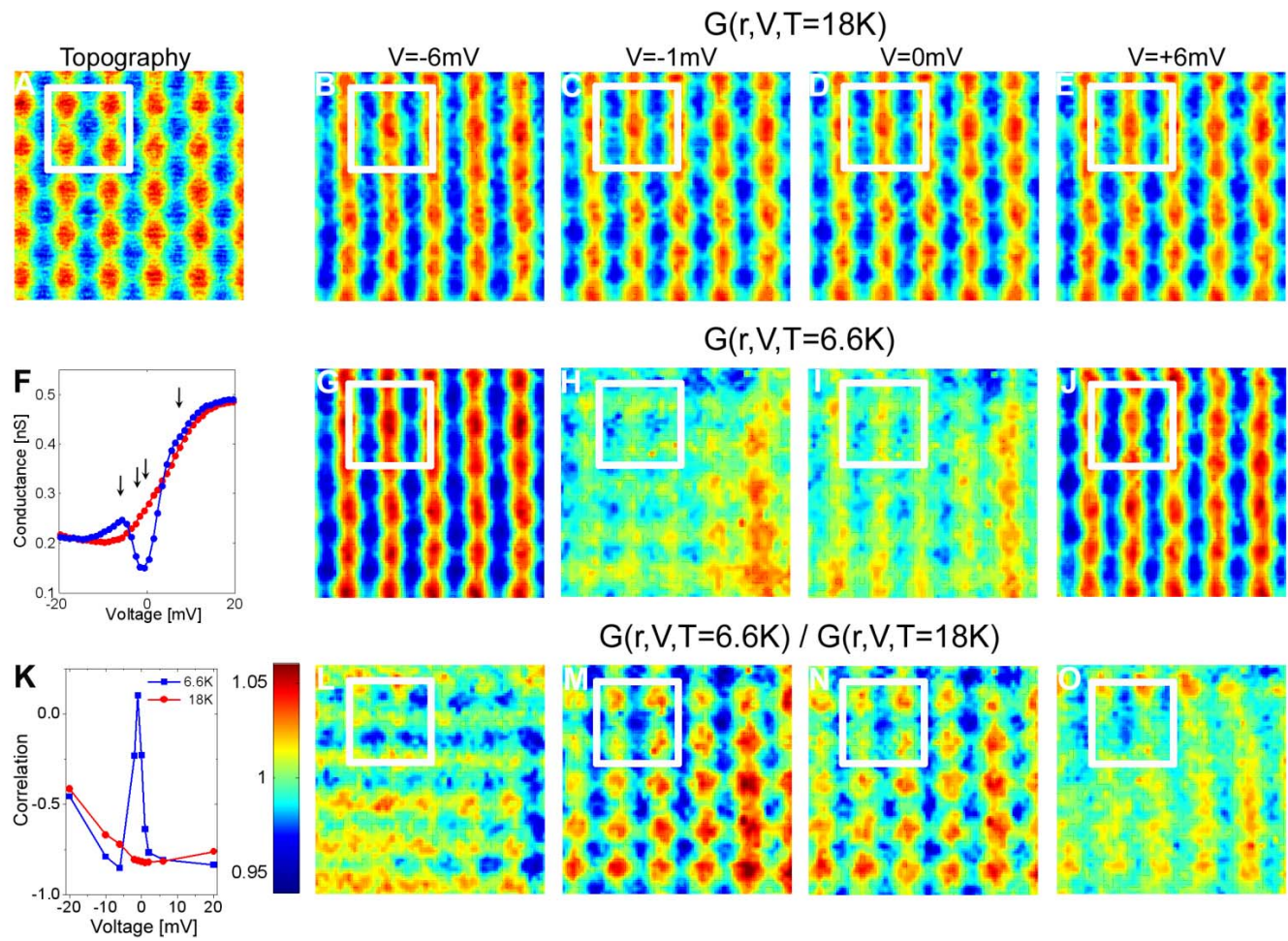

Topography

G(r,V,T=18K)

V=-6mV    V=-1mV    V=0mV    V=+6mV

G(r,V,T=6.6K)

G(r,V,T=6.6K) / G(r,V,T=18K)

**Supplementary Information for "Visualizing the Formation of the Kondo Lattice and the Hidden Order in URu$_2$Si$_2$"**

Pegor Aynajian, Eduardo H. da Silva Neto, Colin V. Parker, Yingkai Huang, Abhay Pasupathy, John Mydosh, and Ali Yazdani

We performed the tunneling measurements in a homebuilt variable-temperature scanning tunneling microscope that operates in the temperature range from 6K – 180K. Lower temperature (2 and 4 K) measurements were performed on a low temperature scanning tunneling microscope.

The single crystal URu$_2$Si$_2$ samples used for this study were grown in an optical floating-zone furnace. Small, flat crystals were oriented along the crystallographic axes and cut into sizes suitable for STM measurements (~2 X 2 X 0.2 mm). The samples were cold cleaved on a surface perpendicular to the c-axis at $T$~10K in UHV and transferred in situ to the microscope head. Differential conductance (d$I$/d$V$) measurements were performed using standard lock-in techniques. The relative variation in d$I$/d$V$(V) at a single point does not depend on the height of the tip and is therefore assumed to be representative of the local electronic density of states (DOS) within an overall normalization.

## I. Identifying the cleavage planes

Multiple surfaces have been obtained after cleaving (Fig. S1, Fig.1). In ~55% of the cases, the topography is as shown in Figure S1A. This 300 Å image of the surface (termed surface (A)) displays a square lattice ($a \sim 4.1$ Å) that corresponds to either the U or Si spacing (S1). The Ru layer has a smaller interatomic spacing as it is rotated 45° with respect to the U layer (see Fig.3B). The step size between equivalent surfaces (~ 4.8 Å) corresponds to half the unit cell. In approximately 45% of the cases, the topography is as shown in Figure S1B. The surfaces in this 1000 Å image show a 'cigar-like' reconstruction (inset) with a width of 2 lattice constants (~8.2Å, surface (B)). The step height between the reconstructed surfaces (~ 4.8 Å) is also equivalent to half the unit cell.

Averaged spectroscopic dI/dV measurements on the different surfaces at T=4 K are shown in Fig.S2. The spectra on the reconstructed surfaces do not necessarily resemble that of the bulk.

## II. Fitting the dI/dV spectra

Figure S3 illustrates our fitting procedure for the electronic density of states shown in Fig.2. The high temperature data (above $T_{HO}$) were fitted to the sum of a Fano lineshape and a V-like background (dashed lines in Fig.S3). The background is obtained from a phenomenological fit to the highest temperature data (120 K) consisting of a linear spectrum with different slopes for hole and electron tunneling. The lower temperature data (below $T_{HO}$) were fitted to the same function multiplied by an asymmetric BCS-like gap function (dotted line). The fitting functions were convolved with the Fermi function appropriate for the temperature. The results of the fits are shown as red lines both below and above the hidden order. The extracted hidden order gap magnitude is plotted in Fig.4C.

### III. Temperature evolution of the hidden order gap

To clarify the temperature dependence of the hidden order gap, and show that it evolves more rapidly than simple thermal broadening, we thermally broaden the data below $T_{HO}$ all to 18K. The divided raw and the artificially broadened gaps are shown in Fig. S4. Though the spectra (Fig. S4B) are smeared out due to the artificial broadening, they still show the opening of a gap already visible at 15K. This confirms that the gap evolves with temperature rather than filling up due to thermal broadening.

### IV. Normalization-independence of the onset temperature

We see from the measured spectra that the hidden order gap is not seen in the raw spectra by T=18 K. Accordingly, the spectra were normalized by the 18 K curve. To further verify that our results are independent of this choice of normalization temperature, we normalize the spectra by dividing them with a higher temperature data. Plotted in Fig. S5 are the averaged spectra of Fig. 2B below T=40K divided with the 40K spectrum, which shows that the hidden order gap indeed opens below 18K. Since the Kondo-Fano resonance has somewhat broadened at 40 K relative to 18 K, an additional background is produced by this choice of normalization temperature. However, the hidden order gap can be clearly distinguished from this weak background.

## Fig. S1

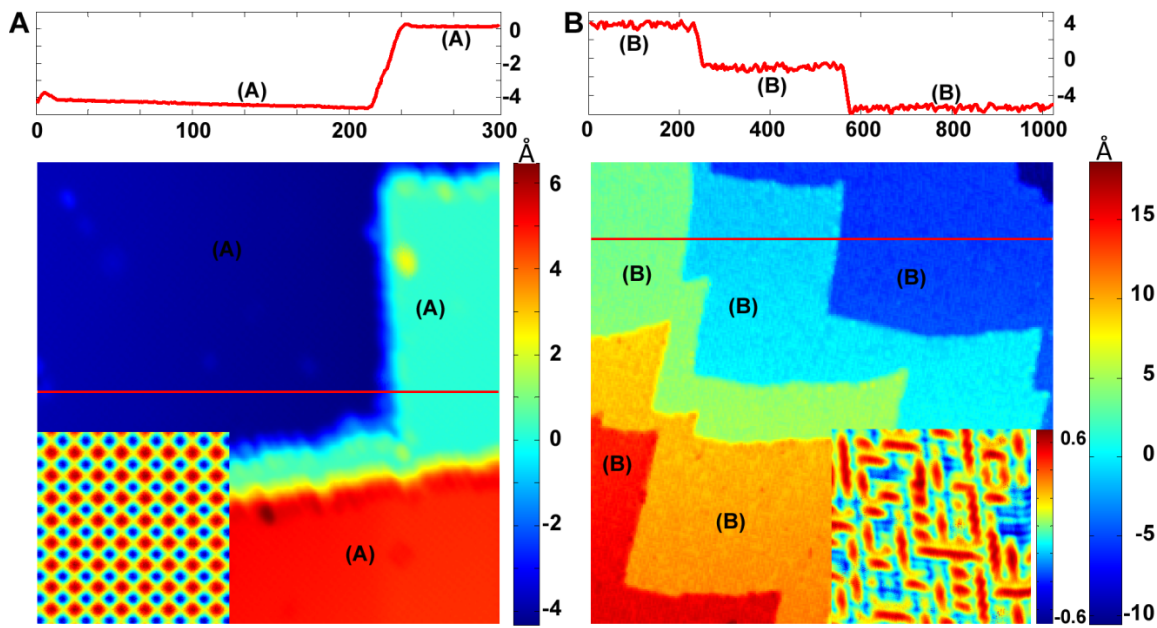

**Figure S1: (A)** Constant current topographic image (-100 mV, 50 pA) over a 300 Å area of the atomic surface (Surface A), showing atomic terraces. The inset shows the magnification of the atomically ordered surface. **(B)** Constant current topographic image (100 mV, 80 pA) over a 1000 Å area of the reconstructed surface (Surface B). The inset shows the magnification of a 100 Å area showing the reconstruction. In both panels step heights are ~4.8 Å, corresponding to half the unit cell. The red lines show the locations of horizontal cuts through the data.

**Fig. S2**

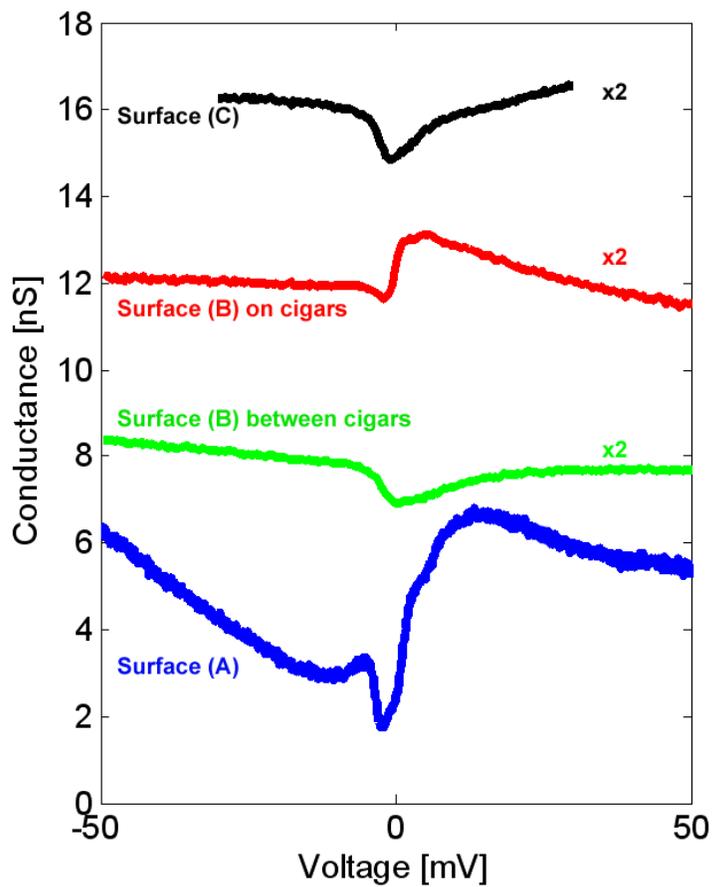

**Figure S2:** Averaged dI/dV measurements on the different observed surfaces. The spectra on Surface B strongly depend on the location. In between the "cigars" the spectra shows a gap-like feature similar to that of Surface C. On top of the "cigars" the spectra shows a resonance-like feature. The spectra are offset by 4nS for clarity.

**Fig. S3**

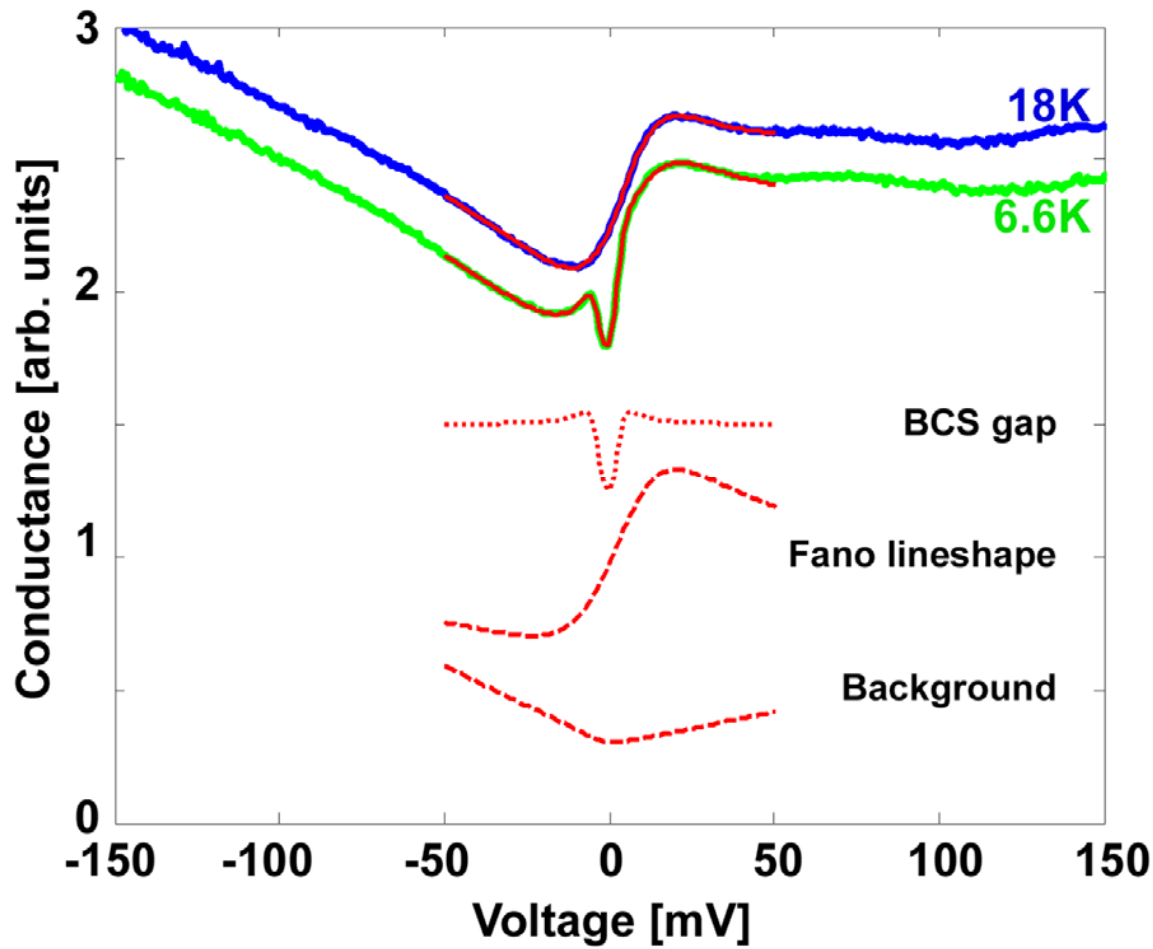

**Figure S3:** Averaged d*I*/d*V* at 18 K (blue) and 6.6 K (green). Above the hidden order temperature, the data are fitted to a Fano lineshape and a V-shaped background (dashed lines). Below the hidden order temperature, the data are fitted with the same function multiplied by a BCS-like gap (dotted line). The red solid lines represent the fit to the data above (sum of dashed lines) and below (sum of dashed lines multiplied by the dotted line) the hidden order temperature.

**Fig.S4**

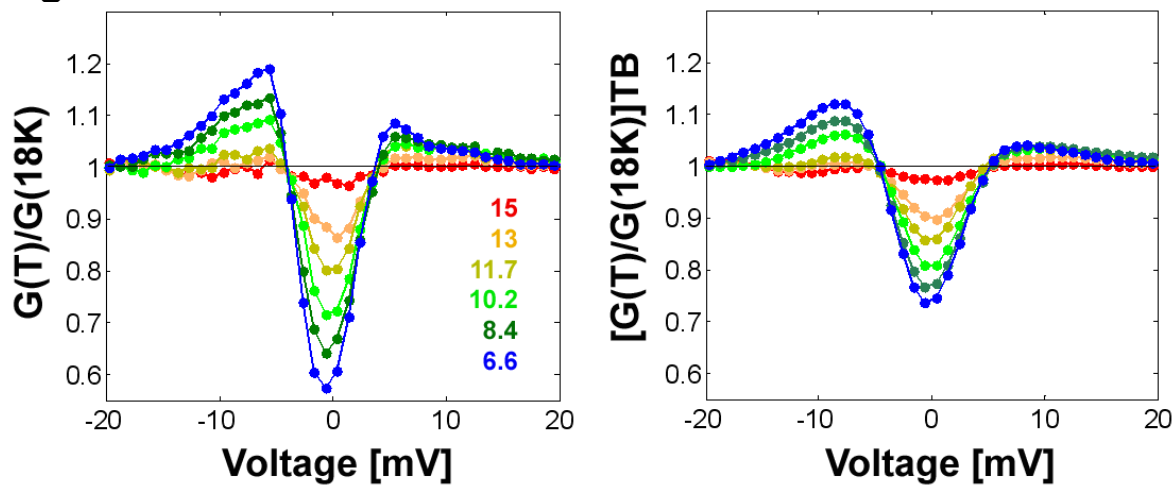

**Figure S4:** (**A**) The experimental data below $T_{HO}$ divided by the 18 K data showing the evolution of a gap in the DOS. (**B**) The same data as in (**A**) all thermally broadened to 18K which clearly demonstrates that the gap evolves with temperature more rapidly than simple thermal broadening.

**Fig. S5**

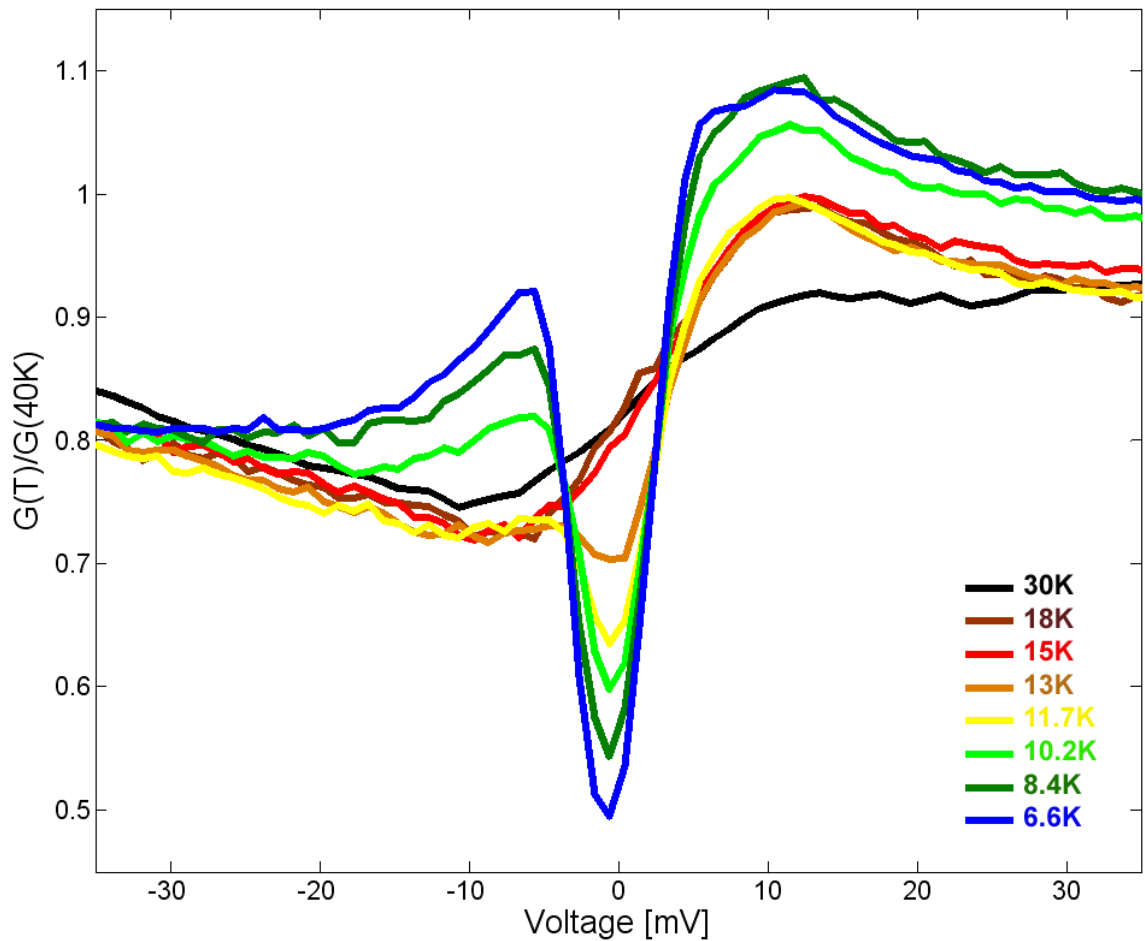

**Figure S5:** The averaged d*I*/d*V* for T<40K divided by the T=40K data. The data shows the development of a gap near the Fermi energy below 18K. The spectra also display the residue of the asymmetric Fano lineshape.